\documentclass[final,3p,times]{elsarticle}

\usepackage{lineno}
\modulolinenumbers[5]

\biboptions{sort&compress}

\usepackage{amsmath,amsthm,amssymb,pifont}
\usepackage{subfig}
\usepackage{caption}
\usepackage[usenames]{color}
\usepackage{anysize}
\marginsize{2.5cm}{2.5cm}{2cm}{2.5cm} 

\newcommand{\Ni}[1]{Ni$^{#1+}$}
\newcommand{\Mn}[1]{Mn$^{#1+}$}
\newcommand{\F}{\ensuremath{\textsf{F}}}
\newcommand{\A}{\ensuremath{\textsf{A}}}
\newcommand{\K}{\ensuremath{\textsf{K}}}

\newcommand{\nn}{\ensuremath{\langle n,m \rangle}}
\newcommand{\nnn}{\ensuremath{\langle\hspace{-0.5mm}\langle n,m\rangle\hspace{-0.5mm}\rangle}}
\newcommand{\Ham}{\ensuremath{\mathcal{H}}}
\newcommand{\Sn}{\ensuremath{\boldsymbol{S}}}
\newcommand{\Tmin}{\ensuremath{\theta_{\text{min}}}}
\newcommand{\opd}[3]{\ensuremath{#1_{_{#2,#3}}^{\,\dagger}}}
\newcommand{\op}[3]{\ensuremath{#1_{_{#2,#3}}}}
\newcommand{\exa}[2]{\ensuremath{\scalebox{1.2}{e}}{\scalebox{1.2}{$^{#1\,#2}$}}}

\begin{document}
\begin{frontmatter}
\title{Magnetic excitations of perovskite rare-earth nickelates: RNiO$_3$}


\author[a,b]{Ivon R. Buitrago\corref{corresp}}
\cortext[corresp]{Corresponding author}
\ead{ivonnebuitrago@cab.cnea.gov.ar}

\author[b,c]{Cecilia I. Ventura}

\address[a]{Instituto Balseiro, Univ. Nac. de Cuyo and CNEA, 8400 Bariloche, Argentina}
\address[b]{(CONICET) Centro At\'omico Bariloche-CNEA, Av. Bustillo 9500, 8400 Bariloche, Argentina}
\address[c]{Universidad Nacional de R\'{\i}o Negro, 8400 Bariloche, Argentina}

\begin{abstract}
 The perovskite nickelates RNiO$_3$ (R: rare-earth) have been studied as potential multiferroic compounds. A certain degree of charge disproportionation in the Ni ions has been confirmed by high resolution synchrotron power diffraction: instead of the nominal \Ni{3} valence, they can have the mixed-valence state \Ni{(3-\delta)} and \Ni{(3+\delta)}, though agreement has not been reached on the precise value of $\delta$ (e.g. for NdNiO$_3$, $\delta=0.0$ and $\delta=0.29$ were reported). Also, the magnetic ground state is not yet clear: collinear and non-collinear Ni-O magnetic structures have been proposed to explain neutron diffraction and soft X-ray resonant sccattering results in these compounds, and more recently a canted antiferromagnetic spin arrangement was proposed on the basis of magnetic susceptibility measurements. This scenario is reminiscent of the situation in the half-doped manganites. 
   
In order to gain insight into the ground state of these compounds, we studied the magnetic excitations of some of the different phases proposed, using a localized spin model. With the purpose of describing the charge disproportionation, we include two kinds of Ni-spins with different magnitudes. As for the magnetic couplings, we include: nearest-neighbor (NN) and next-nearest-neighbor (NNN) Heisenberg-like interactions, respectively for the ferromagnetic and antiferromagnetic couplings present in the collinear phases. To describe the non-collinear phases, and as already proposed for other multiferroics, we also consider NN Dzyaloshinskii-Moriya-type couplings to allow for the possibility of a relative angle $\theta$, between NN spins in the two different magnetic sublattices. Using a simplified spin chain model for these compounds, we first analize  the stability of the collinear, orthogonal, and intermediate phases in the classical case. We then explore the quantum ground state indirectly, calculating the spin excitations  obtained for each phase, using the Holstein-Primakoff transformation and the linear spin-wave approximation. For the collinear and orthogonal ($\theta=\pi/2$) phases, we predict differences in the magnon spectrum which would allow to distinguish between them in future inelastic neutron scattering experiments.
\end{abstract}

\end{frontmatter}

\section{Introduction}
\label{sec.Introduction}
The ferroelectric oxides with magnetic ordering have attracted much attention since they offer the possibility of controlling the electric polarization or the magnetic ordering by applying magnetic or electric fields, respectively,  
a desirable feature in the design of electronic devices~\cite{Cheong}.
 However, finding these multiferroic oxides has not been an easy task. Though there are some of them which have a simultaneous ferroelectric character and magnetic ordering, usually the coupling between these is very weak and therefore poorly controlled with applied fields. 
In 2004  Efremov \emph{et al.}~\cite{Efremov} suggested that in manganites (RMnO$_3$ R: rare-earth), in addition to simultaneous charge and magnetic ordering, a charge disproportionation (CD) of the Mn ions would be needed for these materials to become multiferroic. This CD  means that instead of the nominal valence \Mn{3},  mixed valences \Mn{(3-\delta)} and \Mn{(3+\delta)} should be present. Later, van den Brink and Khomskii~\cite{VanDenBrink} discussed about the possibility of ferroelectricity related to charge disproportionation in rare earth perovskite nickelates of the type RNiO$_3$ (R=rare earth). In fact, in 2000 Mizokawa \emph{et al.}~\cite{Mizokawa} had studied a multiband $d-p$ model for perovskite transition metal oxides, suggesting that it could describe those nickelate compounds, and found an antiferromagnetic ground state with charge ordering centered either in the O-2p orbitals, for relatively large charge-transfer energy (as in PrNiO$_3$ and NdNiO$_3$), or with charge-ordering in the transition metal 3d orbitals, for negative charge transfer energies (relevant for YNiO$_3$).

These nickelates ($R$=rare-earth, or Y) present a metal-insulator transition~\cite{Lacorre,Torrance,Medarde} at temperature T$_\text{MI}$, and antiferromagnetic ordering below the Neel temperature $T_\text{N} (\leqslant T_\text{MI})$ with a possible ordering of \Ni{(3-\delta)} and \Ni{(3+\delta)} ions~\cite{Garcia1992,Garcia1994,Alonso1999,Alonso2000,Fernandez2001,Scagnoli2006,Medarde2008,Munoz2009,Garcia2009,Alonso2013}, with various values of the charge disproportionation $\delta$ as discussed below. As one example, in NdNiO$_{3}$ it was found that $T_\text{MI}=T_\text{N}=200$ K~\cite{Lacorre,Garcia1992}. 

In 2009, Giovanetti \emph{et al}~\cite{Giovannetti} showed by first principles calculations, that in nickelates simultaneous charge and magnetic ordering could be present, as well as a charge disproportionation of the Ni ions, and electrical polarization would thus be induced. In their work, they calculated the electrical polarizations obtained for three of the magnetic phases previously proposed for nickelates, along with a specific charge ordering of \Ni{2} and \Ni{4} ions corresponding to a charge disproportionation of $\delta = 1$. The magnitude and direction of the electrical polarization induced would indicate the underlying magnetic order in these oxides, at present still not clear.
 
\begin{figure}[!hbt]
  \centering
\includegraphics[width=0.4\columnwidth]{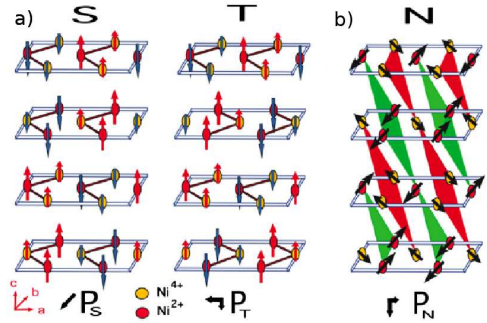}
 \caption{(Reproduced from Ref~\cite{Giovannetti}-Fig.1.) Schematic view of the charge and magnetic structures of RNiO$_3$ oxides. (a) Collinear up-up-down-down magnetic structure~\cite{Garcia1992}, (b) Non collinear magnetic structure~\cite{Scagnoli2006}}
 \label{fig.1}
\end{figure}

As shown in Figure~\ref{fig.1}, the magnetic orderings which they studied~\cite{Giovannetti} are: i) the {\it\textbf{S-collinear phase}} first proposed by Garcia \emph{et al.}~\cite{Garcia1992} in X-ray (XRD) and neutron diffraction (ND) experiments for PrNiO$_3$ and NdNiO$_3$, and later by Fern\'andez \emph{et al.}~\cite{Fernandez2001} for HoNiO$_3$; ii) the {\it\textbf{T-collinear phase}} proposed by Giovannetti~\cite{Giovannetti}; and the iii) {\it\textbf{N-non-collinear phase}} proposed by Scagnoli \emph{et al.}~\cite{Scagnoli2006,Scagnoli2008} for NdNiO$_3$ based on soft X-ray resonant scattering at the Ni-L$_{2,3}$ and Nd-M edges.

The {\it\textbf{S-collinear phase}} (see Figure~\ref{fig.1}), is characterized \cite{Giovannetti} by a checkerboard charge order of \Ni{(3-\delta)} and \Ni{(3+\delta)} ions, corresponding to spins $S_1$ and $S_2$ respectively, along with a magnetic structure defined by the pro\-pa\-ga\-tion vector \mbox{${\bf k}=(1/2,0,1/2)$}, not seen in other perovskite oxides. This involves alternating ferromagnetic (FM) and antiferromagnetic (AF) couplings along the three pseudocubic axes such that every Ni-spin is coupled FM with three of its nearest neighbors (NN) and AF with the remaining ones. Regarding the magnetic cell, this structure can be pictured as formed by $ab$ planes stacked in $c$ direction in the form A$^+$A$^+$A$^-$A$^-$, where in A$^-$ all spins are inverted with respect to A$^+$. Notice that on each plane, there are FM zigzag chains along $b$, which are coupled AF to each other. Experimentally, the direction of the moments within each plane appears to be either along $a$ (Ref.~\cite{Garcia1992}), or in the $ac$ plane (Refs.~\cite{Alonso1999,Fernandez2001}). Notice that the {\it\textbf{T-collinear phase}}, differs from the {\it\textbf{S-phase}} in the stacking of the zig-zag chains between adjacent planes: in the {\it\bf{S phase}} all zig-zag chains point in the same direction, whereas in the {\it\bf{T phase}} in alternate planes they point in opposite directions~\cite{Giovannetti}. The {\it\textbf{N-non-collinear phase}} has the same charge order as both collinear phases considered~\cite{Giovannetti}. However, its magnetic structure corresponds to a spin spiral, in which the spins in FM planes perpendicular to the [101] direction appear rotated around the [010]-axis between consecutive planes. Note that this {\it\textbf{N-non-collinear phase}} is different from other non-collinear phases proposed for nickelates: in Ref.~\cite{Fernandez2001}, these planes are alternatively FM and AF, while in Ref.~\cite{Munoz2009} the FM $ab$ planes are stacked and rotated $\theta\approx 76^\circ$ along [001]. Apart from these phases, recently a canted antiferromagnetic spin arrangement was suggested on the basis of magnetic susceptibility measurements~\cite{Kumar2013}.

A wide set of values has been reported for the charge disproportionation $\delta$  found in  different rare-earth nickelates, as we describe next. For PrNiO$_3$ and NdNiO$_3$, the first studied compounds, $\delta = 0$ according to Refs.~\cite{Garcia1992,Garcia1994}. However, more recently $\delta \sim 0.21$ was reported for PrNiO$_3$~\cite{Medarde2008}, whereas for NdNiO$_3$ a value of $\delta\sim 0.29$ was estimated in Ref.~\cite{Garcia2009} while \Ni{(2.5\pm\delta')} states with $\delta' \sim 0.16$ follow from Ref.~\cite{Scagnoli2006}. For YNiO$_3$ in Ref~\cite{Alonso1999} $\delta\sim 0.28$ was estimated, which coincides with the value in the study through the whole series of $R=$Y, Ho, Er, Tm, Yb, Lu in Ref.~\cite{Alonso2000}, where  $\delta=$0.28, 0.38, 0.32, 0.36, 0.33, 0.33, were respectively reported. For HoNiO$_3$, nevertheless, a larger value $\delta\sim0.48$ can be estimated from the reported magnetic moments in Ref~\cite{Fernandez2001}. For TmNiO$_3$ and YbNiO$_3$, from isomeric shifts in Ref.~\cite{Alonso2013}, $\delta\sim0.14$, and $0.16$ respectively, were estimated, values which correspond to approximately half the indicated CD in Ref~\cite{Alonso2000}. For DyNiO$_3$ in Ref.~\cite{Munoz2009} $\delta\sim0.52$ is found for the non-collinear phase which best agrees with their experiments. 

In the present work, as a first approach to the study of the problem in nickelates, we study the magnetic excitations of a one-dimensional (1D) chain, like the ones included in the collinear and non-collinear phases analized in Ref.~\cite{Giovannetti}. We used the localized spin model to be presented in next section, where the possibility of charge disproportionation is included by considering Ni-spins with eventually different magnitudes. Regarding magnetic couplings, in our model we include the minimal set required to describe the collinear as well as the non-collinear phases proposed.
That is, FM nearest-neighbor (NN) and AF next-nearest-neighbor (NNN) Heisenberg-like interactions, and to describe non-collinear phases also a NN Dzyaloshinskii-Moriya-type (DM) coupling~\cite{Dzyaloshinsky,Moriya}, to allow for the possibility of a relative angle $\theta$ between NN spins in different magnetic sublattices. 
We analize the stability of the collinear, orthogonal, and intermediate phases in the classical model. Then, we explore the quantum ground state indirectly, 
by calculating the spin excitations obtained for each phase, showing that  for the collinear and orthogonal ($\theta=\pi/2$) phases proposed, differences in the magnon spectrums are to be expected, which would allow to distinguish between them experimentally.

\section{Simplified chain model: generic intermediate phase}
\label{sec.Model}

\begin{figure}[!h]
  \centering
  \subfloat[\label{fig.2a}]{
    \includegraphics[width=0.25\linewidth]{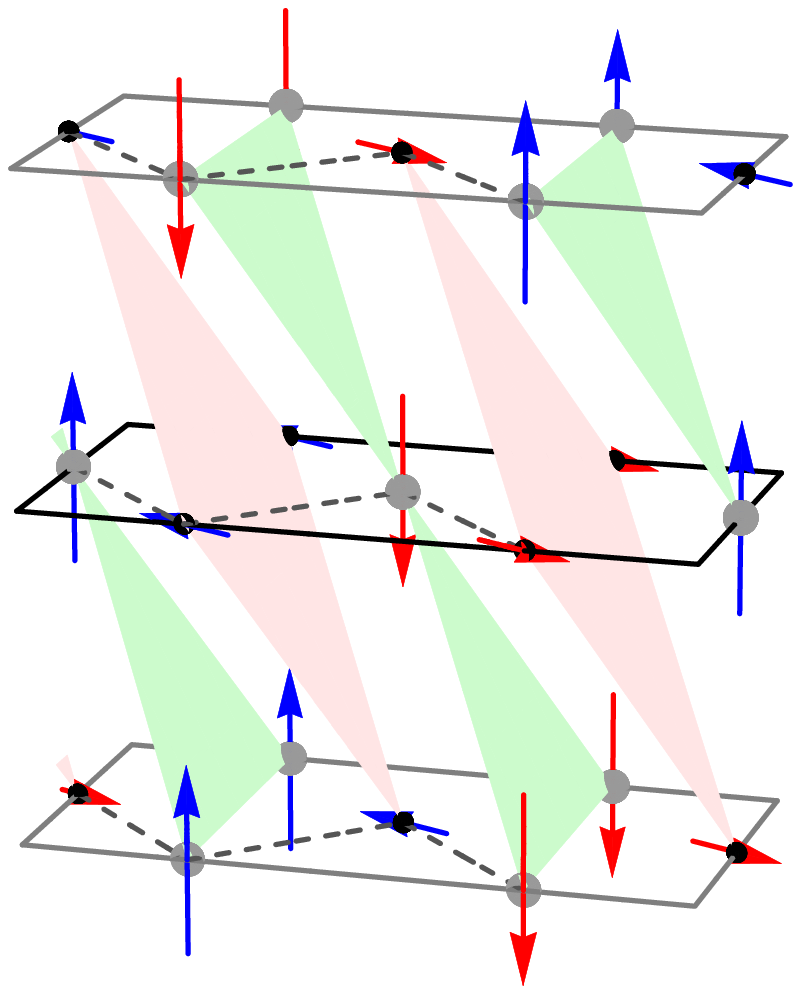}
  }\hspace{1cm}
  \subfloat[\label{fig.2b}]{
    \includegraphics[width=0.5\linewidth]{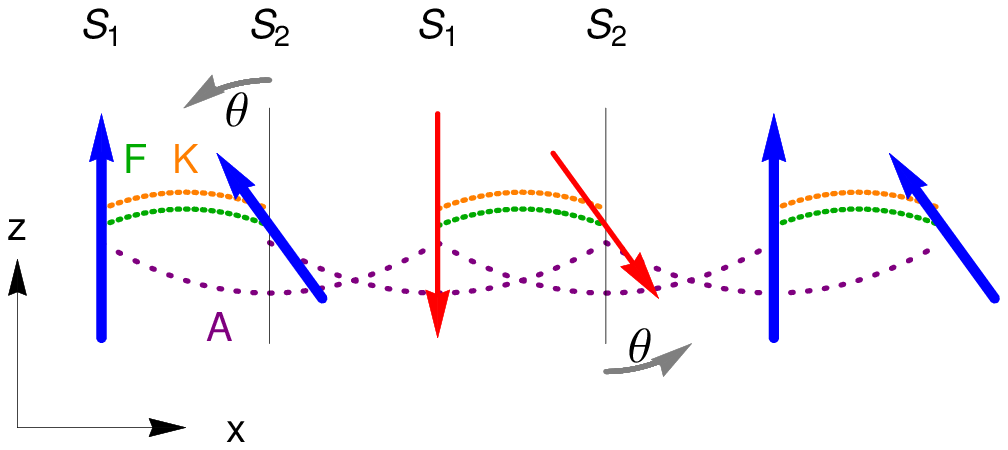}
  }
  \caption{(a) Schematic view of the charge and magnetic orderings under discussion for RNiO$_3$ oxides. With dashed lines we represent the spin chains present, object of our present study. The charge disproportionation \Ni{(3-\delta)} and \Ni{(3+\delta)} is represented, respectively, by gray and black circles. (b) Generic intermediate phase, for one spin chain: i.e. a spin wave, with spins of alternate magnitude. Here the $S_1$-sublattice is considered fixed, while the $S_2$-sublatttice is rotated by an angle $\theta$ with respect to the former.
  }
  \label{fig.2}
\end{figure}

In order to describe the main ground state phases proposed for nickelates mentioned in previous section, we study a simplified chain model and propose a generic phase, which we call the ``intermediate phase''  which, as respective limiting cases, can describe the collinear and the orthogonal phases. 
As a first approach, here the three-dimensional nickelate compounds are studied using a simplified model: representing them by chains of localized spins, shown by dashed lines in Figure~\ref{fig.2a}. 

In the localized spin model we consider two kinds of Ni-spins with different magnitudes in order to describe the charge disproportionation. As shown in Figure~\ref{fig.2b}, and to take into account the phases proposed in experiments~\cite{Garcia1992,Alonso1999,Fernandez2001,Scagnoli2006}, along the chains we consider a unit cell composed by four spins: two of them with magnitude $S_1$ representing the \Ni{(3-\delta)} sublattice, and two other ones with magnitude $S_2$ for \Ni{(3+\delta)}, being $\delta$ a measure of the Ni charge disproportionation (CD). The main difference between the experimentally proposed collinear and orthogonal phases is the relative orientation between the two antiferromagnetic sublattices, which we describe by angle $\theta$ characterizing the intermediate phase, shown in Figure~\ref{fig.2b}. Notice that the sign of $\theta$ will determine the ``helicity´´ of the spin chain. The collinear phase\cite{Garcia1992,Alonso1999,Fernandez2001,Giovannetti} is characterized by $\theta=0$, while the orthogonal phase\cite{Scagnoli2006,Giovannetti} corresponds to $\theta=\pi/2$ and all chains have equal helicity. It is worth mentioning that in Ref.~\cite{Fernandez2001} a slightly different non-collinear phase is proposed: with $\theta\sim 0.44\pi$, and anisotropic helicity (along the z-direction, alternating ``helicity´´ is proposed for consecutive chains). 

Regarding magnetic couplings, in our model we include: FM nearest-neighbor (NN) and AF next-nearest-neighbor (NNN) Heisenberg-like interactions, to describe the collinear phases. To describe non-collinear phases, the model also includes a NN Dzyaloshinskii-Moriya-type (DM) coupling~\cite{Dzyaloshinsky,Moriya}, like the one previously used in Ref.~\cite{Koshibae} to analize the spin excitations in the distorted NiO$_2$ planes in La$_2$NiO$_4$, in order to allow for the possibility of a relative angle $\theta$ between NN spins in the two different magnetic sublattices. 
We found that the minimal model of localized spins which could describe the single chains present in the two limiting phases experimentally proposed~\cite{Garcia1992,Scagnoli2006} as well as the generic intermediate phase with other $\theta$ values as in Ref.~\cite{Fernandez2001}, requires the inclusion of those three magnetic couplings. In particular, if we picture the 4-spin unit cell as formed by two ``dimers'' (plotted in different color/linewidth in Figure~\ref{fig.2b}), an ``intra-dimer'' ferromagnetic (FM) coupling $\F$ and a Dzyaloshinskii-Moriya (DM) like coupling $\K$ are required, as well as a NNN antiferromagnetic (AF) coupling $\A$ in each magnetic sublattice (see Figure~\ref{fig.2b}). 

With the above considerations, we studied the following spin chain Hamiltonian:
\begin{gather}
    \Ham=-\F\sum_{\substack{\nn/\in D}}\Sn_n\cdot\Sn_m^{'} +     \K\sum_{\substack{\nn/\in D}}\widehat{\boldsymbol{y}}\cdot(\Sn_n\times\Sn_m^{'}) 
    +\A\sum_{\substack{\nnn}} \Sn_n\cdot\Sn_m 
    \label{ec.hamiltonian}
\end{gather}
where $D$ indicates spins inside dimers, $\nn$ or $\nnn$ indicate nearest-neighbor (NN) or next-nearest-neighbor (NNN) spins, respectively. Here, all couplings are considered positive, 
and the sign of $\K$ determines the helicity of the spin chain: $\K>0$ produces a counterclockwise rotation like in Figure~\ref{fig.2b}, while $\K<0$ would induce a clockwise rotation. The primes refer to the use of $\theta$-rotated local spin quantization axes for $S_2$-spins, as detailed later. 

\section{Results and discussion.}
\label{sec.results}
\subsection{Phase stability in the classical model}\label{sec.PS-classic}
\quad\\[-1mm]
As a first step, we analize the classical behaviour of the spin chain model of Equation~\eqref{ec.hamiltonian}. The energy of the generic intermediate phase characterized by angle $\theta$, per unit cell, for the case of classical spins is obtained as:
\begin{gather}
  E_{I} (\theta)=-2\,S_1\,S_2\Big(\F\cos\theta + \K\sin\theta\Big)    -\A\left(S_1^2+S_2^2\right)
  \label{eq.Eclassical}
\end{gather}
in terms of the magnetic coupling parameters $\F$, $\K$, and $\A$.

\begin{figure}[!h]
  \centering
  \subfloat{
    \includegraphics[width=0.7\linewidth]{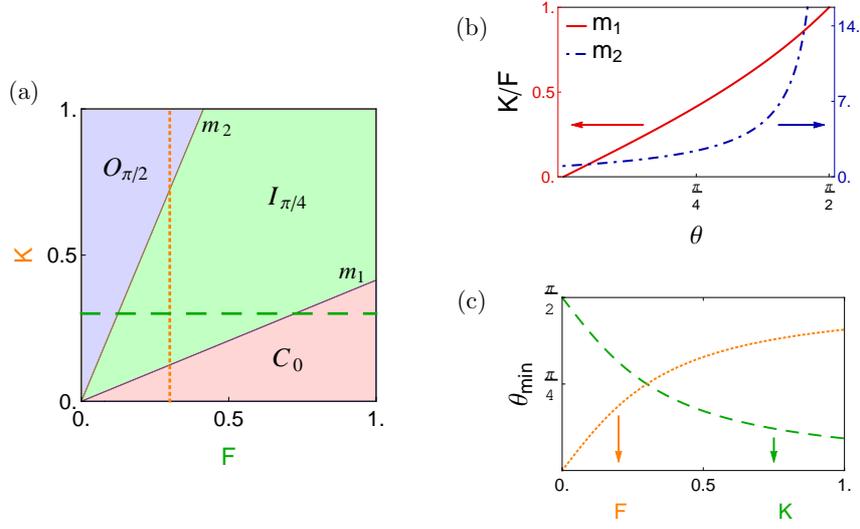}}
  \caption{Analysis of stability of the different phases in the classical model. Parameters: $\A=1$; $S_1=S_2=0.5$. (a)~Classical phase diagram including: the $\theta=\pi/4$ intermediate phase, the collinear phase ($\theta=0$) and the orthogonal ($\theta=\pi/2$) phase. (b) Angular dependence of the slopes of phase boundaries $m_1$ and $m_2$ given by Eq.\eqref{slopes}. (c) $\Tmin$ as a function of $\F$, for $\K=0.3$ (dashed line); $\Tmin$ as a function of $\K$, for $\F=0.3$ (dotted line).}
  \label{fig.3}
\end{figure}

Figure~\ref{fig.3}a shows a phase diagram in ($\F,\K$) space, obtained by comparing the classical energies of three phases: the collinear ($\theta=0$), the orthogonal ($\theta=\pi/2$) and the $\theta=\pi/4$ intermediate phases for $A=1$ and equal spin magnitudes $S_1 = S_2$. 

To explore the dependence on angle $\theta$, one can also compare analytically the energies given by Eq.\eqref{eq.Eclassical} for a generic $\theta$-angle intermediate phase, with the energies of the collinear and the orthogonal phases. The two boundaries for the three observed regions in ($\F,\K$) space (like shown in Figure~\ref{fig.3}a, for $\theta=\pi/4$) are found to be given by linear functions, with respective $\theta$-dependent slopes as shown in Figure~\ref{fig.3}b :  
\begin{gather}
  m_1=\frac{1-\cos\theta}{\sin\theta} \qquad m_2=\frac{\cos\theta}{1-\sin\theta}
  \label{slopes}
\end{gather}
independent of the sublattice spin magnitudes. Thus, fixing the angles of the phases included, the phase diagram in ($\F,\K$) space would not be modified even if different spin values were used, as the energies of the different phases are rescaled proportionally. In Figure\ref{fig.3}b the angular dependence of $m_1$ and $m_2$ is exhibited. 

From \eqref{eq.Eclassical} one can also obtain the angle $\Tmin$ which leads to the intermediate phase with minimum classical energy, for any set of coupling parameters:
\begin{gather}
  \frac{dE_{I}(\theta)}{d\theta}=0   \quad \Rightarrow  \quad  \Tmin=\arctan\left(\frac{\K}{\F}\right).
\end{gather}
Notice that $\Tmin$ is independent of NNN coupling parameter $\A$, and only depends on the ``intra-dimerÂ´Â´ coupling ratio: $\K$/$\F$. In Figure~\ref{fig.3}c we plot $\Tmin$ along specific lines in parameter space, marked in Figure~\ref{fig.3}a. We show the monotonously decreasing $\Tmin$ as a function of $\F$, for $\K=0.3$ and, as could be expected, confirm that for $\F=0$ the orthogonal phase $\theta=\pi/2$ represents the stable ground state. The monotonous increase of $\Tmin$ with $\K$, for $\F=0.3$, is also shown, and we here confirm that for $\K=0$ the collinear phase $\theta=0$ is stable.
Notice also, in Figure~\ref{fig.3}c, that the $\theta=\pi/4$ intermediate phase will only be stable when $\K=\F$. 

\begin{figure}[!h]
  \centering
  \subfloat[Dependence on $\F$, at $\K=0.5$.]{\label{fig.4a}
    \includegraphics[width=0.45\linewidth]{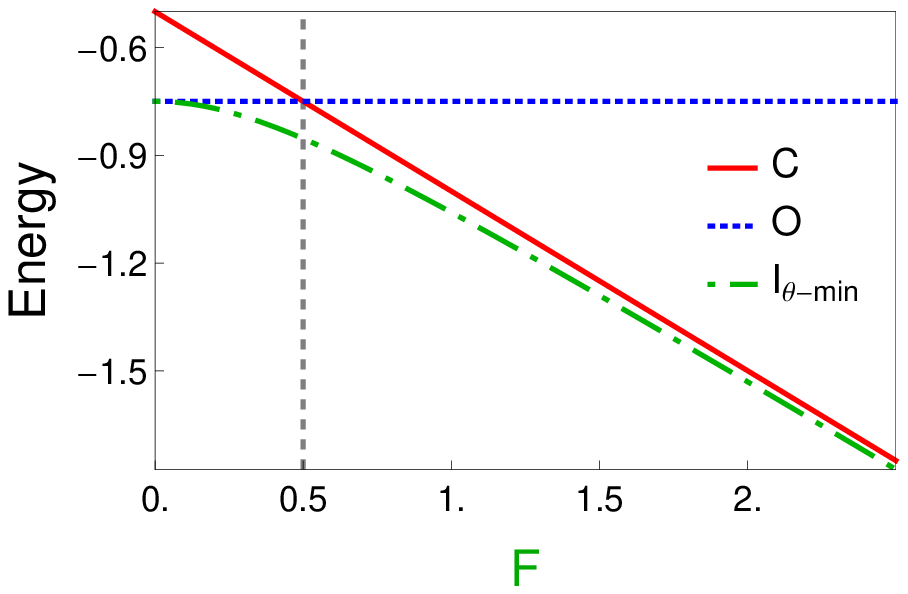}}
  \hspace{1cm}
  \subfloat[Dependence on $\K$, at $\F=0.5$.]{\label{fig.4b}
    \includegraphics[width=0.45\linewidth]{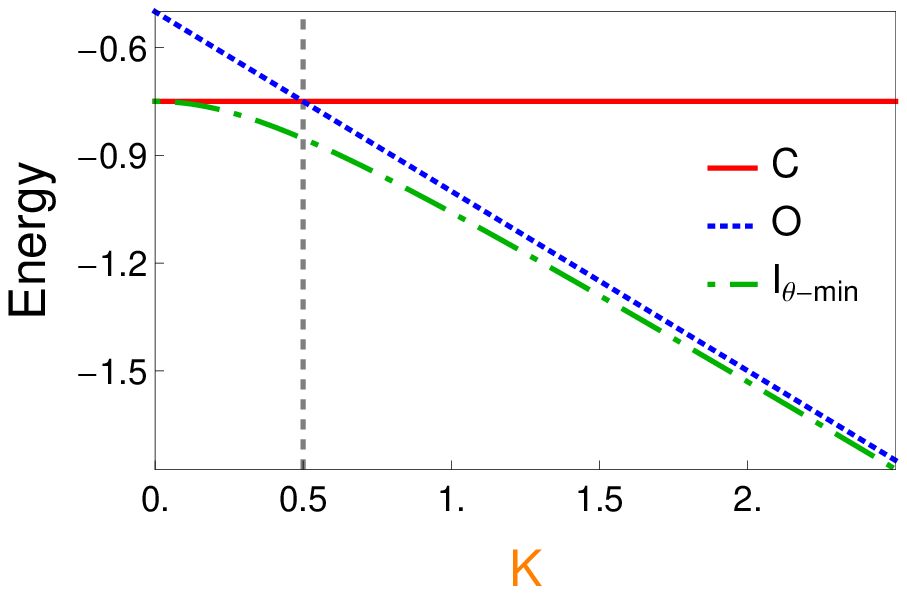}}
  \caption{Classical energy of the: collinear, orthogonal and $\theta=\theta_{min}$ intermediate phases as a function of coupling parameters $\F$ in (a), and $\K$ in (b) Other parameters: $\A=1$; $S_1=S_2=0.5$.}
  \label{fig.4}
\end{figure}

Finally, in Figs.~\ref{fig.4a} and \ref{fig.4b} the classical energies of the collinear phase, the orthogonal phase and the $\theta= \Tmin$ intermediate phase, are plotted as functions of $\F$ and $\K$. It becomes clear that, with our simplified model, for each ($\F$, $\K$) set of parameters, one intermediate phase (the one with $\theta = \Tmin$) is always the classically stable ground state, merging with the proposed collinear and orthogonal phases in the appropriate limits. 

\pagebreak
\subsection{Calculation of the quantum spin excitations}
\label{sec.calculations}
\quad\\[-1mm]
To calculate the quantum magnons of our chain model for nickelates at low temperatures, given by Hamiltonian~\eqref{ec.hamiltonian}, we start by performing a local rotation of the $S_2$ sublattice spin quantization axes by an angle $\theta$, with respect to the $S_1$ sublattice, using the following transformation ($S_n^y=S_n^{y'}$): 
\begin{gather}
    S_n^x= \cos\theta\,S_n^{x'} - \sin\theta\,S_1^{z'}\qquad S_n^z= \sin\theta\,S_n^{x'} + \cos\theta\,S_1^{z'}
\end{gather}

Next, with the Holstein-Primakoff transformation the Hamiltonian is rewritten in terms of bosonic operators, and the Linear Spin Wave approximation (LSW) is used:
\begin{gather}
  \begin{split}
    \begin{aligned}
      S_n^x&=\sqrt{\frac{S_n}{2}}\,\big(a_n^{\dagger}+a_n\big) \\[2mm]
      S_n^x&=\sqrt{\frac{S_n}{2}}\,\big(b_n^{\dagger}+b_n\big) 
    \end{aligned}
    \qquad
    \begin{aligned} 
      S_n^y&=i\sqrt{\frac{S_n}{2}}\,\big(a_n^{\dagger}-a_n\big) \\[2mm]      
      S_n^y&=-i\sqrt{\frac{S_n}{2}}\,\big(b_n^{\dagger}-b_n\big)
    \end{aligned}
    \qquad 
    \begin{aligned}
      S_n^z&=S_n-a_n^{\dagger}a_n\\[2mm]
      S_n^z&=-S_n+b_n^{\dagger}b_n
    \end{aligned}
  \end{split}
\end{gather}
where ($a^{\dagger},a$) operators refer to the spin up sublattice, and ($b^{\dagger},b$) to the spin down sublattice. Though the resulting Hamiltonian includes one, two and three operator terms, in LSW we consider only those written in terms of two operators. One operator terms change the energy of ground state. Introducing the Fourier transform of the boson operators, we obtain the following Hamiltonian:
\begin{align}
  \begin{split}
    \Ham=\sum_q \bigg\{&\lambda_1\,\left(\opd{a}{1}{q}\op{a}{1}{q} + \opd{b}{3}{q}\op{b}{3}{q}\right) + \lambda_2\,\left(\opd{a}{2}{q}\op{a}{2}{q} + \opd{b}{4}{q}\op{b}{4}{q}\right) +\delta_+\,\left(\opd{a}{1}{q}\op{a}{2}{q}+\op{a}{1}{q}\opd{a}{2}{q} + \opd{b}{3}{q}\op{b}{4}{q}+\op{b}{3}{q}\opd{b}{4}{q}\right)\\[2mm]
    &+\delta_-\,\left(\opd{a}{1}{q}\opd{a}{2}{-q}+\op{a}{1}{q}\op{a}{2}{-q} + \opd{b}{3}{q}\opd{b}{4}{-q}+\op{b}{3}{q}\op{b}{4}{-q}\right) + \gamma_1^*(q)\,\opd{a}{1}{q}\opd{b}{3}{q} + \gamma_1(q)\,\op{a}{1}{q}\op{b}{3}{q} \\[2mm]
    &+ \gamma_2^*(q)\,\opd{a}{2}{q}\opd{b}{4}{q} + \gamma_2(q)\,\op{a}{2}{q}\op{b}{4}{q}\bigg\}
  \end{split}
  \label{ec.FinalH}
\end{align}
with the coefficients defined as \Big($S=\sqrt{{S_1\,S_2}\phantom{^2}}$\,\Big):
\begin{gather}
  \begin{split}
    \lambda_{1,2}=S_{2,1}\,\Big(\F\cos\theta + \K\sin\theta\Big) + 2\A\,S_{1,2}  \quad  \delta_{\scalebox{1.2}{$\pm$}}=-\frac{S}{2}\Big\{\F(\cos\theta \pm 1) + \K\sin\theta\Big\}  \quad
    \gamma_{1,2}(q)=\A\,S_{1,2}\,\Big(1+\exa{i}{qa}\Big) 
  \end{split}
  \label{ec.coefficientsH}
\end{gather}
Finally, by paraunitary diagonalization of the Hamilonian, we determine the energies of the magnon excitations.
 
\subsubsection{Magnon predictions for the different phases}
\label{sec.Magnons}
\begin{figure}[!h]
  \centering 
  \subfloat[Points of ($\F,\K$) space, where the predicted quantum magnons are shown next. Colour-depth indicates the value of classical angle $\Tmin$, as indicated by the scale included. Unless otherwise stated: $\A=1$, $S_1=S_2=0.5$]{\label{fig.5a}
    \includegraphics[width=0.35\linewidth]{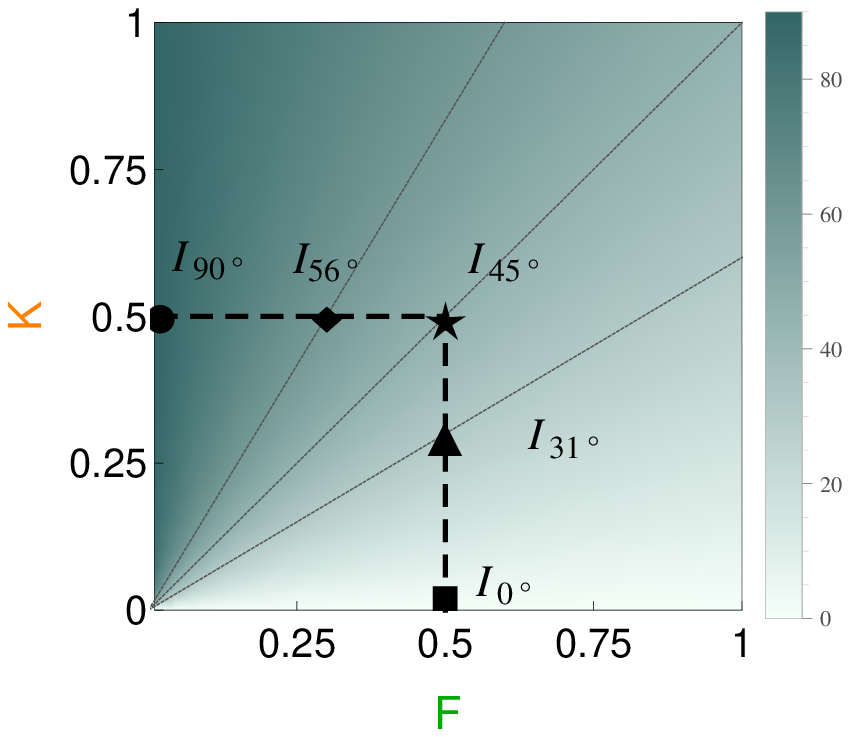}}
  \hspace{1cm}
  \subfloat[Magnons obtained with intra-dimer couplings: \mbox{$\F=0.5$}, $\K=0$ ({\tiny $\blacksquare$} in Fig.~\ref{fig.5a}), for different phases. Solid line: collinear phase ($\theta=0=\Tmin$). Dashed line: orthogonal phase ($\theta=\pi/2$). Dotted lines: collinear phase ($\theta=0=\Tmin$) for $S_1=0.6$ and $S_2=0.4$.]{\label{fig.5b}
    \includegraphics[width=0.4\linewidth]{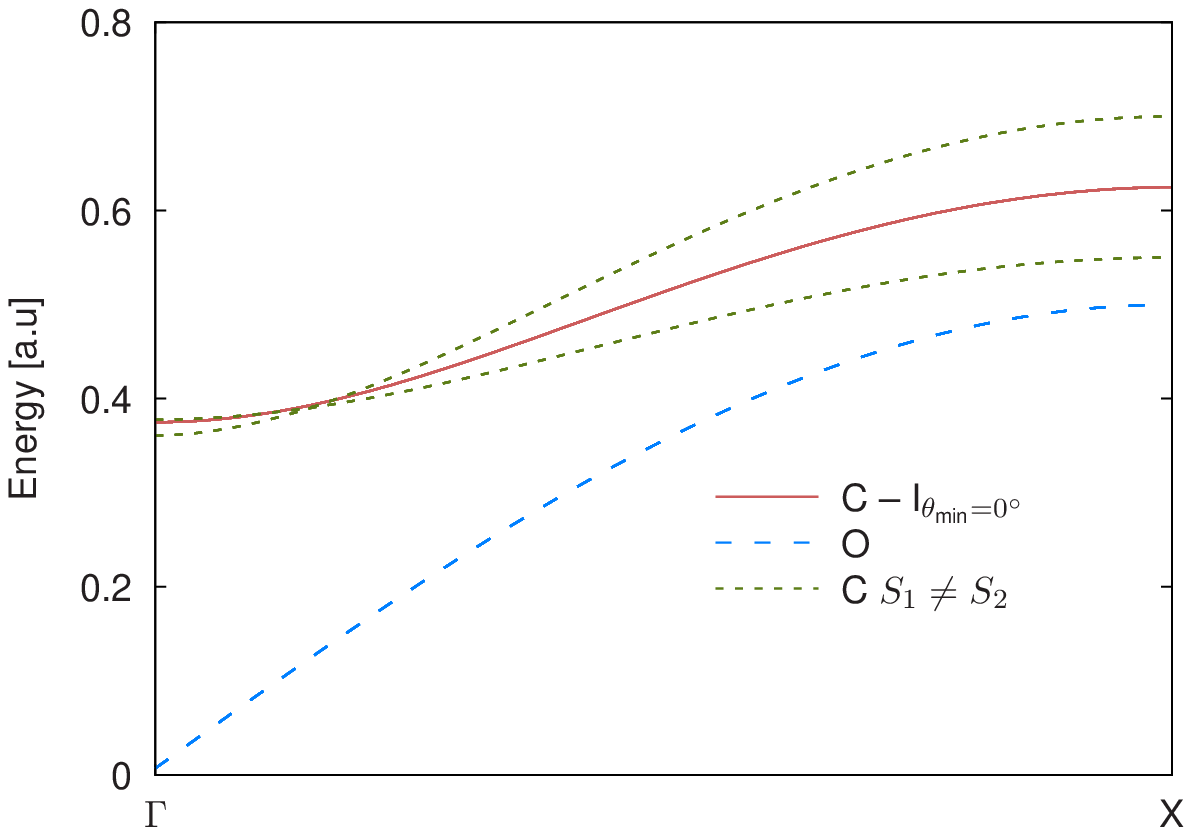}}
  \\
  \subfloat[Magnons corresponding to: \mbox{$\F=0.5$}, $\K=0.3$ ($\blacktriangle$ in Fig.\ref{fig.5a}). Here, the intermediate phase that minimizes the classical energy corresponds to angle: $\Tmin=30.9^{\circ}$.]{\label{fig.5c}
    \includegraphics[width=0.4\linewidth]{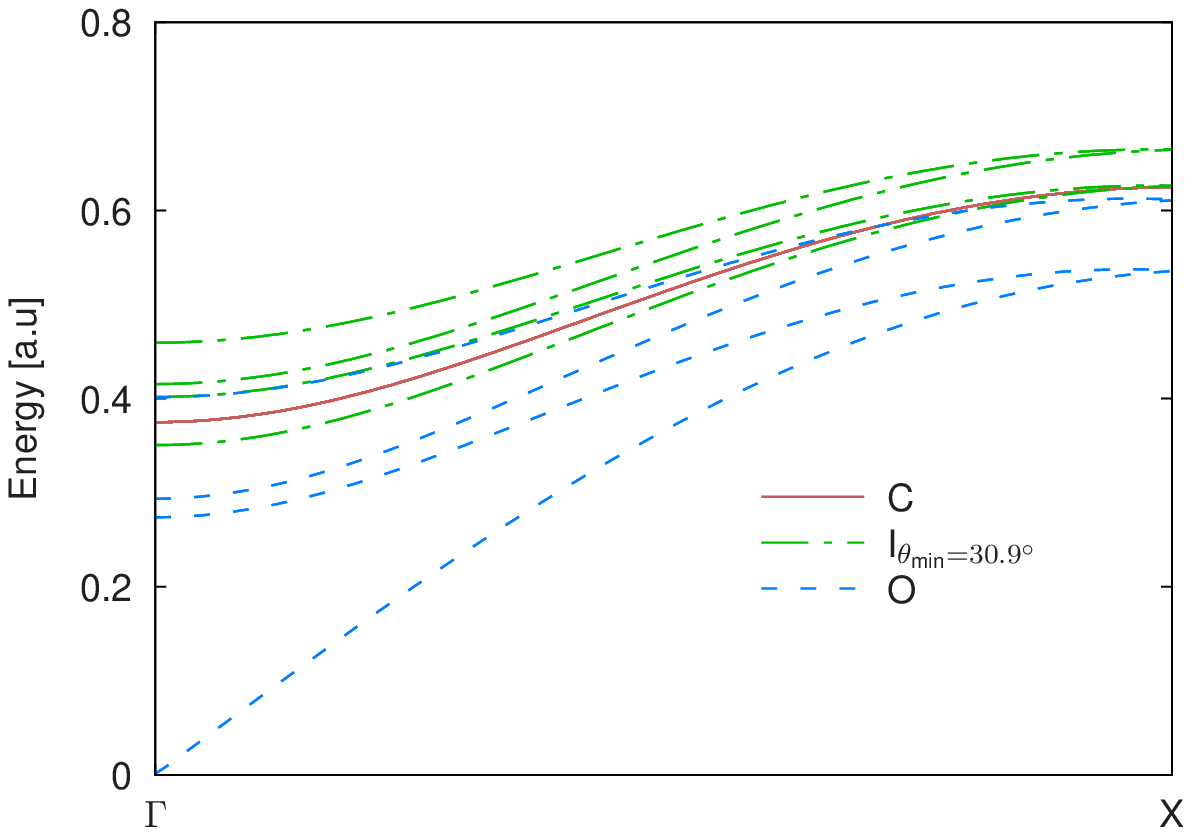}}
  \hspace{1cm}
  \subfloat[Magnons corresponding to: \mbox{$\F=0.5$}, $\K=0.5$ ($\bigstar$  in Fig.\ref{fig.5a}). Here: $\Tmin=45^{\circ}$.]{\label{fig.5d}
    \includegraphics[width=0.4\linewidth]{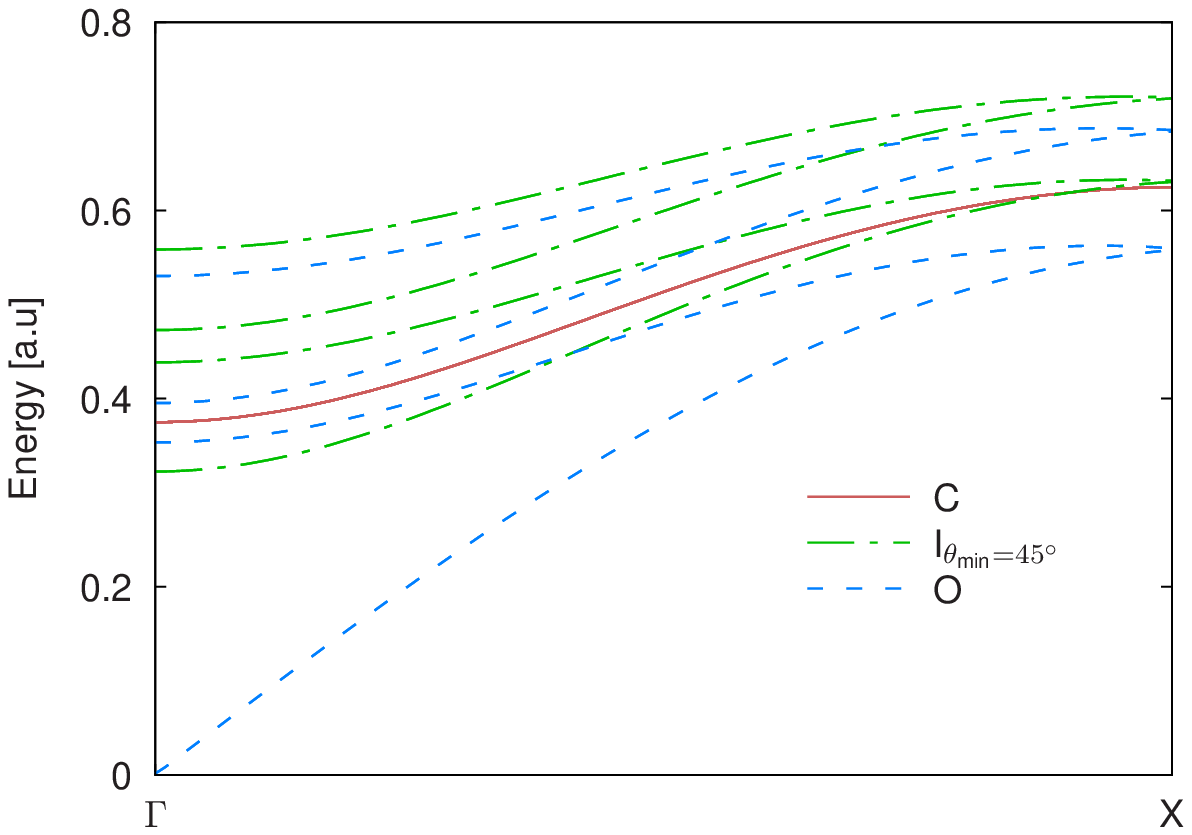}}
  \\
  \subfloat[Magnons corresponding to: \mbox{$\F=0.3$}, $\K=0.5$ (\ding{117}  in Fig.\ref{fig.5a}). Here: $\Tmin=59^{\circ}$.]{\label{fig.5e}
    \includegraphics[width=0.4\linewidth]{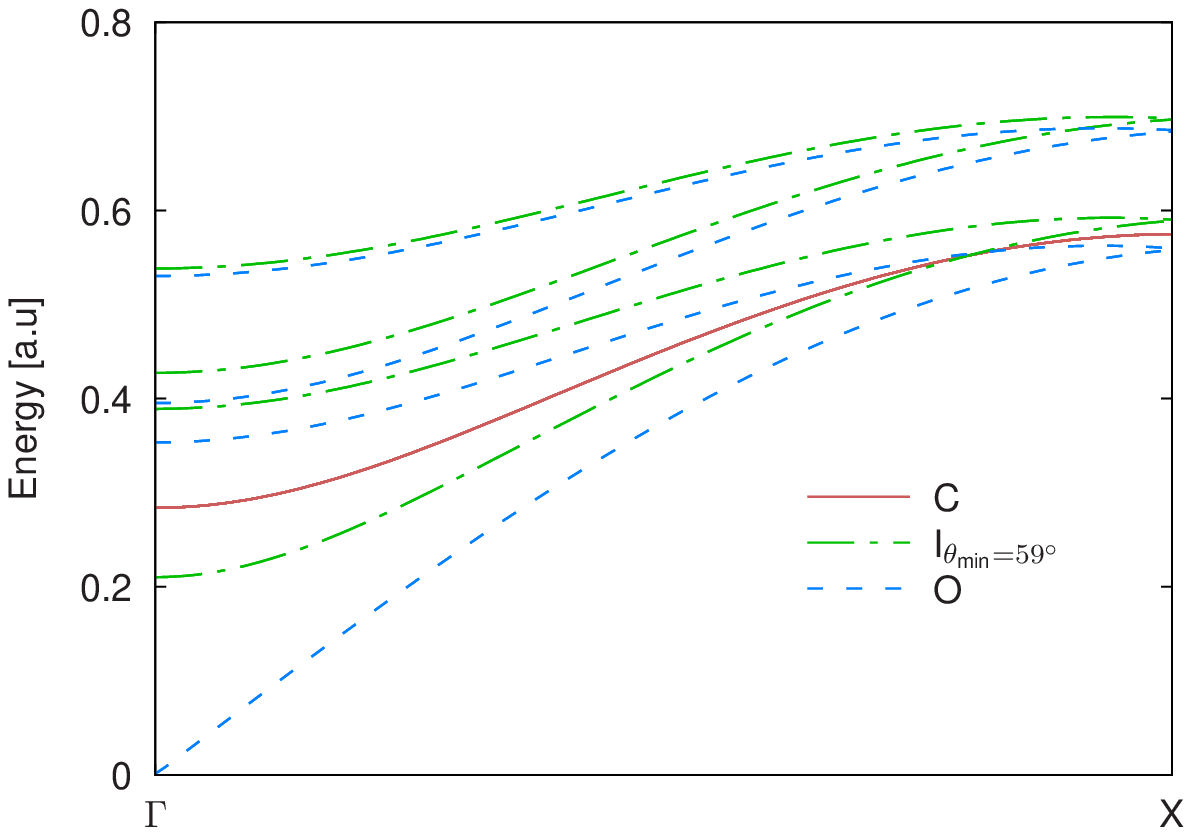}}
  \hspace{1cm}
  \subfloat[Magnons corresponding to: \mbox{$\F=0$}, $\K=0.5$ ({\scriptsize\ding{108}} in Fig.\ref{fig.5a}). Orthogonal phase ($\theta=\pi/2=\Tmin$): dashed lines: for $S_1=S_2=0.5$; dotted lines: for $S_1=0.6$ and $S_2=0.4$.]{\label{fig.5f}
    \includegraphics[width=0.4\linewidth]{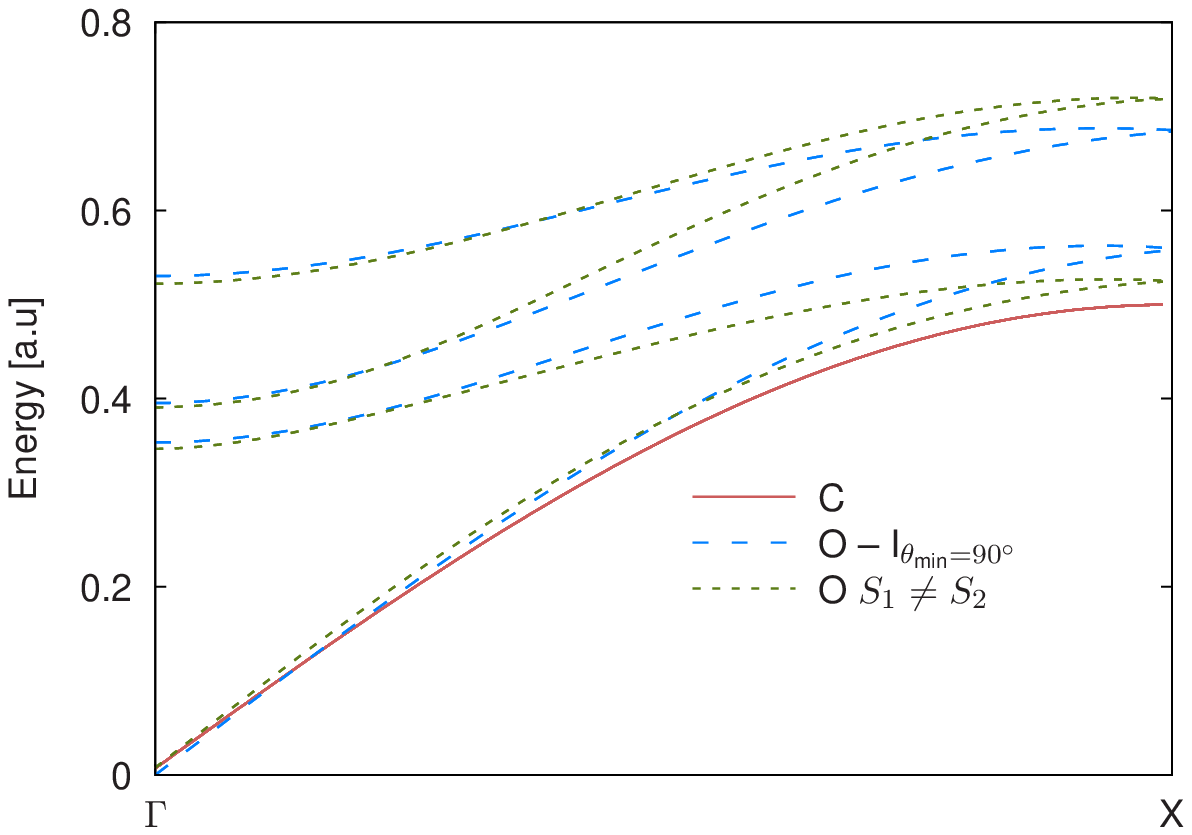}}
  \caption{Magnons for RNiO$_3$ nickelates obtained with our spin chain model for different parameters. Inter-dimer AFM coupling $\A=1$; spins: $S_1=S_2=0.5$, other parameters as detailed in subcaptions. Unless otherwise stated: collinear phase with $\theta=0$ (solid line); orthogonal phase with $\theta=\pi/2$ (dashed lines); $\Tmin$-intermediate phase (dot-dashed lines).}
  \label{fig.5}
\end{figure}

\quad\\[-1mm]
In the following, we discuss the quantum magnons obtained for Hamiltonian \eqref{ec.FinalH} using different pairs of ($\F,\K$) couplings, as detailed in figure~\ref{fig.5a}. The AFM coupling between ``dimers'', $\A=1$, was considered as unit of energy, and except for Fig.~\ref{fig.5b} the spin magnitudes where chosen as $S_1=S_2=0.5$, which corresponds to $\delta=0$. 

In all cases we obtained 16 eigenbands, corresponding to the 16$\times$16 Hamiltonian matrix, of which only the 8 positive branches describe magnon excitations.

First, notice that with the different pairs of ($\F,\K$) couplings used, in Figs.~\ref{fig.5b}-\ref{fig.5f} we are exhibiting the quantum magnons 
which correspond to the intermediate phase with lowest classical energy corresponding to angles $\Tmin$ ranging from 0 to $\pi/2$.  
          
Now, notice that the lowest magnon branch both in Figures \ref{fig.5b} and \ref{fig.5f} is identical to the spin excitation of an antiferromagnetic chain. In Fig.~\ref{fig.5f}, the lowest branch corresponds to the collinear phase ($\theta=0$) with parameters: $\F=0$, $\K=0.5$, hence in~\eqref{ec.coefficientsH} the effect of $\K$ on the magnons disappears, due to the $\sin\theta$ factor. Thus, the only non-vanishing coupling term in the Hamiltonian is determined by $\A$, hence the antiferromagnetic chain excitation obtained. In contrast, in Fig.~\ref{fig.5b} the lowest branch corresponds to the orthogonal phase ($\theta=\pi/2$) with parameters: $\F=0.5$, $\K=0$. In this case, in Eqs.~\eqref{ec.coefficientsH} the effect of $\F$ does not disappear as $\K$ in the previous one. Nevertheless we checked that the dispersion of an antiferromagnetic chain results even when $\F\to\infty$. This can be understood because in this limit the system behaves as an antiferromagnetic chain composed by dimers, each of them consisting of two consecutive spins coupled by $\F$.
  
As shown in Figure~\ref{fig.5b}, the magnons obtained for the collinear phase (solid line), which is the classically stable phase, are all degenerate. We checked that these excitations correspond to spin-flips between two NNN inter-dimer spins: i.e. in sites 1 and 3, or 2 and 4 with the notation of Fig.~\ref{fig.2b}. Both of these spin flips involve the same energy cost related to coupling $\A$, thus being degenerate. As one would expect, we find that considering a certain degree of charge disproportionation, i.e. $S_1\ne S_2$ in the model, some degeneracies are broken due to the lower simmetry of the system: we exemplify this in figure~\ref{fig.5b} by including the case $S_1=0.6$, $S_2=0.4$ ($\delta=0.2$), plotted with dot-dashed lines. This effect is largest at the Brillouin zone edge $X$.

Regarding the orthogonal phase, Figure~\ref{fig.5f} shows its excitations for $\F=0$ and $\K=0.5$ ({\scriptsize\ding{108}} in Fig.~\ref{fig.5a}), case in which it  corresponds to the classical stable phase. Notice  that turning on the $\K$ coupling also reduces the symmetry of the system and, as previously mentioned, breaks magnon degeneracies. Though the analysis of the obtained excitation modes  is more complex for this phase, the symmetry breaking of the two highest magnon branches can be understood as follows. The highest energy magnon branch (with energy $\sim 0.5$ at $\Gamma$, in Fig.~\ref{fig.5f}) involves spin-flip excitations between two NN inter-dimer spins in sites 1 and 4, and in sites 2 and 3, involving an energy cost related to $\A$ and $\K$ couplings. A lower energy cost is payed exciting the magnon branch below it 
(with energy $\sim 0.4$ at $\Gamma$, in Fig.~\ref{fig.5f}): which we checked corresponds to spin excitations of two NNN inter-dimer spins (i.e. between spins 1 and 3, and spins 2 and 4), in which case only the $\K$ coupling is affected. 
 
It is also interesting to compare the effect of the charge disproportionation in the two cases exhibited in Figs.~\ref{fig.5b} and \ref{fig.5f}, respectively. While, as discussed above, in the collinear phase of Fig.~\ref{fig.5b} degeneracies are clearly split by $\delta$, we observe that in the orthogonal phase depicted in Fig.~\ref{fig.5f} no new degeneracy splittings appear, in addition to those originated by the presence of the DM coupling $\K$. In fact, the main $q-$dependent effect produced by $\delta$ in the orthogonal phase is to increase the size of the magnon gap between the upper and lower pairs of branches originated by $\K$. According to our results, the different numbers of magnon branches observed might thus be used to distinguish between the collinear and orthogonal phases, and even to quantify the charge disproportionation. 
 
Also, the figures~\ref{fig.5b}-\ref{fig.5f} show that in the parameter ranges considered none of these phases becomes unstable, being all excitation energies positive. Even though this does not allow us to indirectly determine which of the studied phases would represent the ground state of our chain, it is plausible to infer that the quantum ground state should be similar to the intermediate phase that minimizes the classical energy (i.e. $I_{\Tmin}$: plotted with dot-dashed lines in Figure \ref{fig.5}). This might be justified observing that, in figures \ref{fig.5b}-\ref{fig.5f}, the $\Tmin$-intermediate phase has  excitations with higher energy than the other phases, thus it seems more difficult to create excitations and eventually destabilise the $\Tmin$-intermediate phase.


\section{Summary}

A simplified localized spin chain model was proposed to study the generic intermediate phase in nickelates, able to describe a variable charge disproportionation and relative orientation of consecutive spins along the chain.  

The model includes the following magnetic couplings: nearest-neighbor (NN) and next-nearest-neighbor (NNN) Heisenberg-like interactions, respectively for the ferromagnetic and antiferromagnetic couplings present in the collinear phases. 
To describe the non-collinear phases, we also consider a NN Dzyaloshinskii-Moriya\cite{Dzyaloshinsky,Moriya}-type coupling to allow for the possibility of a relative angle between the different magnetic sublattices. 

We studied: (i) the classical stability of the collinear, orthogonal, and intermediate phases, as possible ground states for these compounds; and: (ii) the quantum ground state indirectly, by calculating the spin excitations resulting from each of those phases. 
 
Our classical results show that for each set of NN (intra-dimer) ferromagnetic and DM  magnetic couplings, always an intermediate phase characterized by an angle $\Tmin$, corresponds to the most stable classical ground state. From the measurements of the two Ni spin magnitudes by Fernandez \emph{et al.}~\cite{Fernandez2001}, in terms of our model one can obtain the following estimation for the relative orientation of consecutive spins along the chain, $\theta\sim 80^{circ}$, and  intra-dimer coupling ratio: $\K/\F\sim 6$.
 
Regarding the quantum magnetic excitations: with our simplified model, we predict the spin excitations to be expected for the collinear~\cite{Garcia1992,Fernandez2001} and the orthogonal phases~\cite{Scagnoli2006,Giovannetti} so far proposed for these compounds, as well as those of the intermediate phase.  For the collinear and orthogonal  ($\theta=\pi/2$) phases, we predict differences in the magnon spectrum which would allow to distinguish between them in future inelastic neutron scattering experiments. In particular, the number of magnon branches would differ between these phases, and the charge disproportionation present might also be quantified: either by the number of branches in the collinear phase, or by the size of the magnon gap in the orthogonal phase.  

Our present study represents a first step towards an understanding of the complex three-dimensional ground state of rare-earth nickelates, to enable  
comparison with the results of future inelastic scattering experiments. The latter are especially desirable in these compounds, 
since the available neutron diffraction data could not discriminate between the different ground states proposed.
Material specific ingredients, as well as the different possibilities of three-dimensional stacking should be considered in future research work. Also, if no obvious signs of destabilization of any of these phases are observed in the predicted magnons, a direct study of the quantum phase diagram would be important.

\section*{Acknowledgements}
We are greatly indebted to Prof. D.I.Khomskii for suggesting us the study of this problem, background material, and his valuable comments. We acknowledge financial support by CONICET (PIP grant 0702, and the fellowship awarded to I.R.B.). C.I.V. is a member of Carrera del Investigador Cient\'{\i}fico, CONICET.

\section*{References}
\bibliography{biblio}

\end{document}